\documentclass[11pt,twoside]{article}

\usepackage{asp2006}
\usepackage{fancyhdr,graphicx,caption,rotating,multirow,lscape}
\DeclareGraphicsExtensions{.eps,.eps.gz,.ps,.jpg,.tiff}

\begin{document}

\title{The Sys-Rem Detrending Algorithm: Implementation and Testing}  



\author{T. Mazeh}
\affil{Wise Observatory, Sackler Faculty of Exact Sciences, Tel Aviv University, Tel Aviv 69978, Israel}
\author{O. Tamuz}  
\affil{Wise Observatory,  Sackler Faculty of Exact Sciences, Tel Aviv University, Tel Aviv 69978, Israel}
\author{S. Zucker}  
\affil{Department of Geophysics and Planetary Sciences, Sackler
  Faculty of Exact Sciences, Tel Aviv University, Tel Aviv 69978, Israel}


\begin{abstract} 
Sys-Rem (Tamuz, Mazeh \& Zucker 2005) is a detrending algorithm
designed to remove systematic effects in a large set of lightcurves
obtained by a photometric survey. The algorithm works without any
prior knowledge of the effects, as long as they appear in many stars
of the sample.  This paper presents the basic principles of Sys-Rem and
discusses a parameterization used to determine the number of effects removed.  
We assess the performance of Sys-Rem on simulated transits
injected into WHAT survey data.  This test is proposed as a general
scheme to assess the effectiveness of detrending algorithms.
Application of Sys-Rem to the OGLE dataset demonstrates the power of
the algorithm. We offer a coded implementation of Sys-Rem to the community.
\end{abstract}

\section{Introduction}
Since the discovery that the planet orbiting HD\,209458
\citep{mazeh2000} transits the disk of its host star
\citep{char2000,henry2000}, many photometric searches for transits
have been put into operation \citep[e.g.,][]{horne2003}. However, till
September 2006 the yield of these searches was surprisingly
small. Only the realization that systematic effects and red noise
\citep{pzq2006} are an impediment to transit detection explained why
many searches detected less planets than expected. The work of \citet{pzq2006}
sharpened the need to account for the presence of red noise in the
survey data.  Sys-Rem \citep{tmz2005}, an algorithm to remove
systematic effects in large sets of lightcurves obtained by
photometric surveys, is designed exactly to answer this need. The
algorithm can detect any effect that appears linearly in many
lightcurves obtained by the survey. Recently, Sys-Rem, together with
other detrending algorithms such as TFA \citep{tfa}, have become
standard tools in transit survey lightcurves processing,
contributing already to the recent detection of several transits
\citep{bakos2006,cc2006}. This paper discusses the
implementation of Sys-Rem and suggests a way to assess its performance.
 
Section 2 reviews the principles of Sys-Rem and Section 3 presents our
stopping criterion, a parametrization to determine the number of effects
to remove.  In Section 4 we propose a test to assess the
effectiveness of detrending algorithms and apply it to
Sys-Rem. Section 5 discusses the application of the algorithm to the
OGLE survey. We conclude with some remarks.

\section{The principle of Sys-Rem}
\label{sec:sysrem}
 
We first started to develop our algorithm in an attempt to correct for
atmospheric extinction, with an approach similar to that of
\cite{ks2003}.  We derived the best-fitting airmasses of the different
images and the extinction coefficients of the different stars, without
having any prior information on the stellar colours.  However, the final 
result is a general algorithm to deal with any linear systematic
effects. In some restricted cases, when one can ignore the different
uncertainties of the data points, this algorithm reduces to the
well-known Principal Component Analysis \citep[][Ch. 2]{mh1987}.
However, when the uncertainties of the measurements vary
substantially, as is the case in many photometric surveys, PCA
performs poorly relative to Sys-Rem.

The principles of Sys-Rem can be easily explained using the original
problem we tried to solve.  Colour-dependent atmospheric extinction is
an obvious observational effect that contaminates ground-based
photometric measurements.  This effect depends on stellar colours,
which are not always known. To correct for the atmospheric extinction
one can find the effective colour of each star, which characterizes
its variation as a function of the airmass of the measurements.

Specifically, consider a set of $N$ lightcurves, each of which is
derived from $M$ images. Define the residual of each observation,
$r_{ij}$, to be the average-subtracted stellar magnitude of the $i$-th
star derived from the $j$-th image, taken at the airmass $a_j$.  We
can then define the effective extinction coefficient $c_{i}$ of star
$i$ to be the slope of the best linear fit to the residuals of this
star as a function of the corresponding airmasses, aiming to remove
the product $c_i a_j$ from each $r_{ij}$. In fact, we search for the
best $c_i$ that minimizes the expression
\begin{equation}
S^2_i=\sum_{j}{{\Bigl(r_{ij}-c_ia_j\Bigr)^2}\over{\sigma^2_{ij}}} \ \ ,
\label{eq:S2i}
\end{equation}
where $\sigma_{ij}$ is the uncertainty of $r_{ij}$. Note that the
derivation of each $c_i$ is independent of all the other $c_i$'s, but
does depend on all the $a_j$'s.

The problem can now be turned around. Since atmospheric extinction
might depend not only on the airmass but also on weather conditions,
we can ask ourselves what is the best estimate of the airmass of each
image, given the known effective colour of each star. Thus, we can
look for the $a_j$ that minimizes
\begin{equation}
 S^2_j=\sum_{i}{{\Bigl(r_{ij}-c_ia_j\Bigr)^2}\over{\sigma^2_{ij}}} \ \ , 
\end{equation}
given the previously calculated set
of $\{c_i\}$.
We can now recalculate new best-fitting coefficients, $c_{i}$, for
every star, based on the new $\{a_j\}$, and
continue iteratively. We thus have an iterative process which in
essence searches for the two sets -- $\{\bar c_i\}$ and $\{\bar
a_j\}$, that best account for the atmospheric extinction.

Many simulations have shown that this iterative process converged to
the same $\{\bar a_j\}$ and $\{\bar c_i\}$, no matter what initial
values were used. Therefore, we suggest that the proposed algorithm
can find the most suitable effective airmass of each image and the
extinction coefficient of each star.

The algorithm, in fact, finds the best two sets of
$\{c_i\ ;i=1,N\}$ and $\{a_j\ ;j=1,M\}$ that minimize the {\it global}
expression
\begin{equation}
S^2=\sum_{ij}{{\Bigl(r_{ij}-c_ia_j\Bigr)^2}\over{\sigma^2_{ij}}} \ \ .
\label{eq:S2}
\end{equation}
Therefore, although the alternating 'criss-cross' iteration process
\citep{gz1979} started with the actual airmasses of the
different images, the values of the final set of parameters $\{\bar
a_j\}$ and $\{\bar c_i\}$ are not necessarily related to the true
airmass and extinction coefficient. They are merely the variables by
which the global sum of residuals, $S^2$, varies linearly most
significantly.  They could represent any strong systematic effect that
might be associated, for example, with time, temperature or position
on the CCD. This algorithm finds the systematic effect as long as the
global minimum of $S^2$ is achieved.

Now, suppose the data are affected by a few different systematic
effects, with different $\{c_i\}$ and $\{a_j\}$. Sys-Rem can be
applied repeatedly, until it finds no more {\it significant} linear effects
in the residuals.

\section{The halting problem}
\label{alpha}

Formally, the process of identifying additional 'systematic' effects
in any set of lightcurves can be applied till there is no variation
left in all lightcurves. To prevent such a situation, it is obvious
that Sys-Rem needs a stopping criterion, which will enable it to
remove the strong systematic effects in the data without removing the
signal of the variable stars, the transit signals in particular.

Our stopping criterion is based on a measure of the strength of each effect
in each lightcurve. We therefore define $\beta$ as the
fractional r.m.s.~removed by subtracting the effect from a specific
lightcurve. We assume that a significant effect yields a large
$\beta$. Note that $\beta$ is defined independently for each
lightcurve and each effect. Our stopping mechanism involves choosing
$\beta_{min}$, so that we apply Sys-Rem subtraction only to effects and
lightcurves with $\beta \geq \beta_{min}$.

In order to estimate $\beta_{min}$ we use the value of $\beta$ found
in a set of randomly generated lightcurves of similar noise structure as the
real ones. For each lightcurve in a given dataset we generate a corresponding random
lightcurve with a randomization technique, by which we keep the
stellar intensities but randomly permute the timing on all
measurements. Such a procedure should get rid of any correlated noise
hidden in the original lightcurves, while keeping the level of the
white noise. We then apply Sys-Rem to the entire set of false
lightcurves, find a Sys-Rem effect in this randomized matrix, and
calculate for each false lightcurve its $\beta$ value. We use the
distribution of the $\beta$ values of the false ligthcurves to derive
$\beta_{min}$, which is the $\beta$ value for which a fraction
$\alpha$ of the random lightcurves have smaller values of
$\beta$. Therefore, $\beta_{min}$ is a monotonic increasing function
of $\alpha$. Choosing $\alpha$ of say, 0.9, means removing only
effects that are stronger than 90 percent of all random effects.  In
Section~4 we try various values of $\alpha$ and find that $0.9$ is indeed a
good choice for the WHAT dataset.

\section{Assessing the performance of a detrending algorithm}
\label{sec:test}

We propose here a general scheme to test the performance of detrending
algorithms. The test is performed on simulated data which have been
generated by injecting simulated transit signals into real data. The
test is conducted by applying the detrending algorithm to the
simulated data and then searching for transits. A transit search
applied after an effective detrending algorithm should detect a large
fraction of the injected transits, and should {\it not} yield many
'false positive' transits that had not been injected into the data.

The test proposed here deserves two comments at this stage. The first
is related to the assumption that the real data do not include many
real transits. Even an accurate set of lightcurves can include at most
only very few transits, so the evaluation of the test should not be
thwarted by the presence of real transits. The second comment has to
do with the fact that the test we propose really checks the
effectiveness of the detrending algorithm {\it together} with the
transit detection technique, applied to a specific dataset. Therefore,
this test does not assess the overall performance of a detrending
algorithm, but only its usefulness when applied to a specific dataset
and used in conjunction with a particular transit detection algorithm.

The detection of a transit candidate in a set of lightcurves
is never an absolute result, and each transit candidate is likely to be a 
false positive with some
probability. Therefore, any transit survey inevitably outputs a list
of transit candidates, prioritized according to some statistic. Thus,
we expect an effective detrending algorithm, Sys-Rem or other, to
increase the priority of the real transits in the candidate list,
reducing the number of false positives and making follow-up more
efficient.

To test how well Sys-Rem performs we injected simulated transit
signals into some of the lightcurves of WHAT field no. 236
\citep{what}. We ran the BLS transit detection algorithm \citep{bls}
on all lightcurves, including the ones with no transit signal, and
constructed a candidate list.  We repeated this procedure after
applying Sys-Rem, with three different values of $\alpha$.

\begin{figure}[!ht]
\centering
\includegraphics[angle=0,height=10cm]{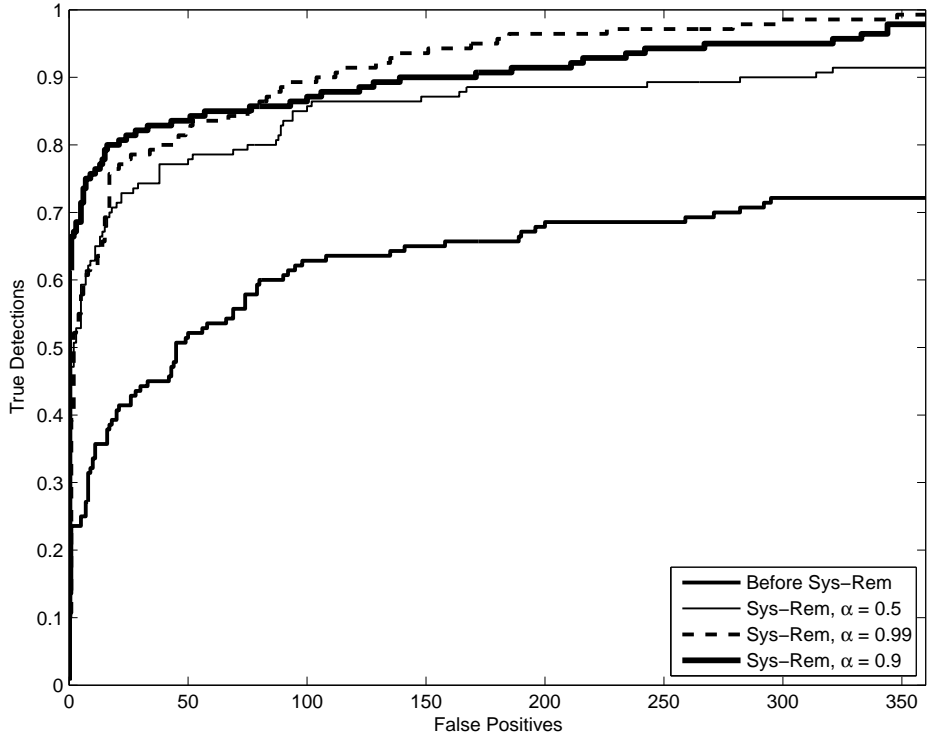}
\caption{
Fraction of transits detected vs. number of false detections, for
different values of $\alpha$.  }
\label{fig:fdr}
\end{figure}

Fig.~\ref{fig:fdr} shows the fraction of injected transits that were
detected by BLS and appear at the top of the candidate list. 
The detection fraction of the injected transits depends on
how far down the list of candidates one goes. Therefore, the figure
presents the detection fraction as a function of the number of false
detections included in the top of the list. In other words, the
figure shows what fraction of the simulated injected transits one
could detect if he is ready to include a given number of false
positive cases.  The higher the graph is, the higher is the probability to
include the transit in the top of the list.

The figure shows that applying Sys-Rem with any of the three values of
$\alpha$ dramatically improved the detectability of the injected
transits. It seems as if for this specific data set and this specific
set of injected transits, Sys-Rem with $\alpha=0.5$ is inferior to the
other two options. Because of the limited observational resources,
most of the present follow-up projects can not allow a high number of
false positive cases, and therefore it seems that Sys-Rem with
$\alpha=0.9$ is the best of the three options tested here.

Note that the number of false positive transits included in the top
of the list depends on the data, its red noise distribution in
particular. Therefore the meaning of the figure is limited to the WHAT
set of lightcurves. 

\section{Application to OGLE}
\label{sec:application}

We applied Sys-Rem to the photometric data collected by the OGLE
survey in three Carina fields: CAR100, CAR104 and CAR105
\citep{udalski2002}.  This dataset includes 1200 measurements of about
one million stars. In each field, we applied Sys-Rem separately to
each of the 8 CCD chips.
 
\begin{figure}[!ht]
\centering
\includegraphics[angle=0,height=10cm]{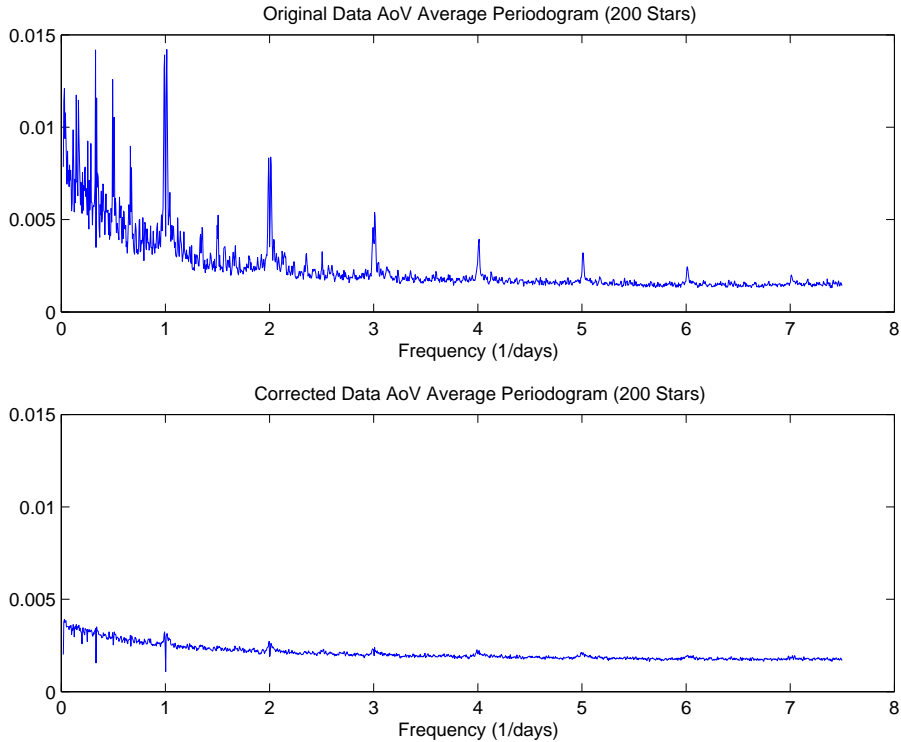}
\caption{
Averaged AoV periodograms of 200 OGLE stars before (upper panel) and
after (bottom panel) applying Sys-Rem}
\label{fig:ogle_periodograms}
\end{figure}

\begin{figure}[!ht]
\centering
\includegraphics[angle=0,height=10cm]{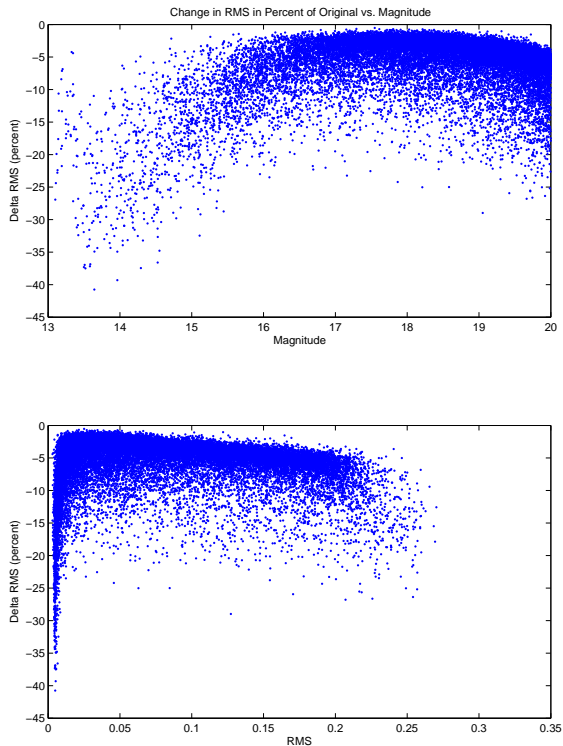}
\caption{
 The reduction in the RMS scatter of the OGLE lightcurves
as a function of the magnitude (upper panel) and the initial
scatter}
\label{fig:deltaRMS}
\end{figure}

 To present the effectiveness of Sys-Rem's application to the OGLE
data we present in Fig.~\ref{fig:ogle_periodograms} and
Fig.~\ref{fig:deltaRMS} some results from the data of chip 8 in field
CAR105. We selected 200 bright stars from the chip, calculated AoV
periodograms (Schwarzenberg-Czerny 1989) for each, and then averaged
the periodograms to produce the upper panel of
Fig.~\ref{fig:ogle_periodograms}.  The averaged periodogram shows
clearly a periodicity of 1 day and its harmonics, and some
low-frequency power. Obviously, this has to do with some systematic
effects.

We generated a similar averaged periodogram after we applied Sys-Rem,
to produce the bottom panel. As is evident, Sys-Rem removes not only
periodic variability with frequencies with integer number of cycles
per day, but also most of the low-frequency variability. Note,
however, that some troughs appear in one day harmonics, which indicate
that Sys-Rem may have removed some true signal along with the
systematics, impairing detection of transits in these
frequencies. \citet{ks2003} got similar results when they searched for
systematics with periodicity of one day and its harmonics.

Fig.~\ref{fig:deltaRMS} shows, for the same chip, the {\it fractional}
change in the RMS scatter obtained by Sys-Rem (in percentage of the
initial scatter), as a function of the magnitude and the original
scatter. The top panel of the figure shows that Sys-Rem is most
effective in reducing the scatter of the brighter stars, where the
systematic noise is more dominant. For those stars the improvement can
get up to 30\% of the original scatter. Note that a small but
substantial improvement can be seen for all stars. The increase of
Sys-Rem improvement for the faint stars is probably due to removal of
systematics associated with background subtraction.

\section{Conclusion}
\label{sec:conclusion}

We have presented a stopping criterion of Sys-Rem, based on the
fraction of the variability subtracted from any lightcurve by removing
the systematic effect. This power is compared with the power
subtracted from random lightcurves with the same noise level. The
comparison is parameterized, so a different threshold of removing
fractional variability can be adopted. We assessed the performance of
Sys-Rem on simulated transits injected into the WHAT survey dataset and
found that Sys-Rem improved the detectability substantially. This is
true for all three values of the stopping parameter used. We propose
this test as a general scheme to assess the effectiveness of
detrending algorithms.

We have presented an application of Sys-Rem to the dataset of the
OGLE transit search. We demonstrated that the algorithm can eliminate
a significant part of the systematic noise hidden in the light
curves. The mainly affected stars are the brighter ones, where the
photon noise is less significant and the systematics grow in
importance.

In a previous conference we \citep{mazeh2006} offered the community an
'overnight cleaning service', through which we apply our Sys-Rem code
to their photometric data and return the data clean of systematics,
ready for search for minute periodic variability. It turned out that
most researches do not like their laundry to be cleaned by
others. Therefore, we are now offering the community to acquire their
own cleaning facility: the code is available in C and can be obtained
upon request from omert@wise.tau.ac.il.

\acknowledgments{We wish to thank A. Udalski for letting us use the
OGLE data.  This work was supported by the German-Israeli Foundation
for Scientific Research and Development}


\end{document}